\documentclass[11pt]{article}
\usepackage{graphicx}

\newcommand{\BABARPubYear}    {03}

\newcommand{\BABARConfNumber} {004}
\newcommand{\SLACPubNumber} {9716}

\input babarsym
\newcommand{\Btaunu}  {\ensuremath{\Bm \to \taum \bar{\nu}\xspace}}
\newcommand{\Blepnu}  {\ensuremath{\Bm \to \ellm \bar{\nu}\xspace}}
\newcommand{\BRBtaunu}{\ensuremath{\BR(\Btaunu)}\xspace}

\newcommand{\Breco}       {\ensuremath{B_{reco}}\xspace}

\long\def\inst#1{\par\nobreak\kern 4pt\nobreak
    {\it #1}\par\vskip 10pt plus 3pt minus 3pt}

\begin{document}
{\pagestyle{empty}
\par\vskip 1cm

\begin{flushright}
\babar-CONF-\BABARPubYear/\BABARConfNumber \\
SLAC-PUB-\SLACPubNumber \\ 

\end{flushright}

\par\vskip 3cm

\begin{center}
\Large \bf  \boldmath A Search for \Btaunu \\
Recoiling Against a Fully Reconstructed \B
\end{center}

\vfill
\begin{center}
\large The \babar\ Collaboration\\
\mbox{ }\\
\today
\end{center}

\vfill

\begin{center}
\large \bf Abstract
\end{center}
 We present a search for the \Btaunu\ decay in a data sample of 82~\invfb 
 collected at the \FourS\ 
 resonance with the \babar\ detector at the SLAC \pep2\ \abf.
 Continuum and combinatorial backgrounds are suppressed by selecting a sample of 
 events with one completely reconstructed \B. The decay products of the 
 other \B\ in the event are analyzed to search for a \Btaunu\ decay.
 The $\tau$ lepton is identified in the following decay channels:
 \taumtoe, \taumtomu, \taumtopi, $\taum \to \pim \piz \nu$, $\taum
 \to \pim \pip \pim \nu$.
 We find no evidence for a signal and set a $90\%$ C.L.
 upper limit of $\BRBtaunu < 7.7 \times 10^{-4}$. 
 We combine this result with another \babar\ measurement searching
 for \Btaunu\ decays in a sample with one \B\ meson reconstructed
 in semi-leptonic channels.
 The two samples are statistically independent. We obtain a
 combined 90\% C.L. upper limit of $\BRBtaunu < 4.1 \times 10^{-4}$.
 All results are preliminary.

\vfill

\begin{center}
Presented at the XXXVIII$^{th}$ Rencontres de Moriond on\\
Electroweak Interactions and Unified Theories, \\
3/15---3/22/2003, Les Arcs, Savoie, France
\end{center}

\vspace{1.0cm}
\begin{center}
{\em Stanford Linear Accelerator Center, Stanford University, 
Stanford, CA 94309} \\ \vspace{0.1cm}\hrule\vspace{0.1cm}
Work supported in part by Department of Energy contract DE-AC03-76SF00515.
\end{center}

\newpage
} 
\begin{center}
\small

The \babar\ Collaboration,
\bigskip

%
B.~Aubert,
R.~Barate,
D.~Boutigny,
J.-M.~Gaillard,
A.~Hicheur,
Y.~Karyotakis,
J.~P.~Lees,
P.~Robbe,
V.~Tisserand,
A.~Zghiche
\inst{Laboratoire de Physique des Particules, F-74941 Annecy-le-Vieux, France }
% ...

\end{center}\newpage

\section{Introduction}
\label{sec:Introduction}
The study of the leptonic decay \Blepnu~\footnote[1]{charge-conjugate 
modes are implied throughout the paper.} is of particular interest because
it is sensitive to the product of the Cabibbo-Kobayashi-Maskawa 
matrix element \Vub\ and the \B\ meson decay constant \fsubb, which
describes the overlap of the quark wave functions inside the \B\ meson
and is only known from theory~\cite{ref:fbtheory}.
The knowledge of \fsubb\ is essential for the extraction of the
Cabibbo-Kobayashi-Maskawa matrix element $|V_{td}|$ from
processes of \BzBzb\ mixing in which the oscillation
frequency is proportional to $f^2_B$.
In the Standard Model the amplitude of the \Blepnu\ decay is due to the 
annihilation of the \b\ and \ubar\ quarks into a virtual \W\ boson.
The resulting expression for the branching fraction is:
\begin{equation}
  \BR(\Blepnu) = \frac{G^2_F m_B}{8 \pi} m_l^2 
        \left( 1 - \frac{m^2_l}{m^2_B} \right)^2 f^2_B \Vub ^2 \tau_B \:, 
  \label{eq:brsm}
\end{equation}
where $G_F$ is the Fermi coupling constant, $m_l$ and $m_B$ are the charged
lepton and \B\ meson masses, $\tau_B$ is the \Bm\ lifetime. 
The dependence of $\BR(\Blepnu)$ on the lepton mass arises from 
helicity conservation, which strongly suppresses the muon and 
electron channels.
Using data from Ref.~\cite{ref:pdg2002}, the Standard Model
expectation in Eq.~(\ref{eq:brsm})
for the $\tau$ channel becomes: 
\begin{equation}
  \BR(\Btaunu) = ( 7.5 \times 10^{-5} ) \frac{\tau_B}{1.674 \ps} 
  \left(\frac{\fsubb}{198 \mev}\right)^2
  \left|\frac{V_{ub}}{0.0036}\right|^2 \: .
  \label{eq:brsmtau}
\end{equation}

While the theoretical dependence of the branching fraction from
the relevant parameters, $\tau_B$, $\fsubb$ and $V_{ub}$,
is straightforward, a search for the \Btaunu\ decay is experimentally 
challenging due to the presence of additional undetectable 
neutrinos in the final state coming from the decay of the $\tau$.

No observation of a \Btaunu\ signal has been reported yet 
in the literature.
The most stringent upper limit has been achieved by the L3
Collaboration~\cite{ref:l3}: 
$\BR(\Btaunu) < 5.7 \times 10^{-4}$ at 90\% C.L.

\section{The \babar\ detector and dataset}
\label{sec:babar}
The data sample used in this analysis was recorded 
at the \FourS\ resonance in 1999-2002
with the \babar\ detector at the \pep2\ asymmetric-energy \epem\ collider 
at the Stanford Linear Accelerator Center.
The integrated luminosity at the center of mass energies near
\FourS\ is 81.9~\invfb, corresponding to 88.9 million \BB\ pairs. 
We also used a Monte Carlo simulation of \BpBm\ generic events with
an equivalent luminosity of 136.9~\invfb and of \tautau\ events 
with an equivalent luminosity of 127.1~\invfb.

The \babar\ detector is described elsewhere~\cite{ref:babar}.
Detection of charged particles and measurement of their momenta 
are performed using a combination of 
a five-layer silicon vertex tracker (SVT) and
a 40-layer drift chamber (DCH) in a 1.5~T solenoidal magnetic field.
A detector of internally-reflected Cherenkov radiation (DIRC) with a 
quartz bar radiator provides charged particle identification.
A finely-segmented CsI(Tl) electromagnetic calorimeter (EMC) is used to detect 
photons  and to identify electrons.
The magnetic flux return system (IFR), which is instrumented
with multiple layers of resistive plate chambers, 
provides muon and long-lived neutral hadron identification.

Charged kaons are identified from the observed pattern of 
Cherenkov light in the DIRC and from the 
d$E$/d$x$ measurements in the SVT and DCH.
Electron candidates are selected according to the ratio of 
EMC energy to track momentum, the EMC cluster shape, the d$E$/d$x$
in the DCH and the DIRC Cherenkov angle, if available.
Muon candidates are selected according to the difference 
between the expected and measured thickness of absorber traversed,
the match of the hits in the IFR with the extrapolated
track, the average and spread in the number of hits per IFR layer,
and the energy deposited in the EMC.

\section{Analysis method}
\label{sec:Analysis}

We select a sample of events with one \B\ meson (\Breco)
completely reconstructed in a variety of hadronic decay modes.
All the tracks and photon candidates in the event
not used to reconstruct the \Breco\ are associated to the other
\B\ meson (recoil \B) and are studied to search for a \Btaunu\ signal.

The advantage of having a sample of fully reconstructed
\Breco\ mesons is to provide a clean environment of \BpBm\ events
with a strong suppression of the combinatorial and continuum backgrounds. 
The drawback is a reduction of the data sample due to the low 
reconstruction efficiency.

\subsection{Fully reconstructed \B\ sample}

The \Breco\ is reconstructed in a set of hadronic modes
that can be summarized as $\B^+ \to D^{(\ast)0} X^+$,
where $D^{(\ast)0}$ is a charmed meson and 
$X^+$ is a system of charged and neutral hadrons 
composed by
$n_1 \pipm + n_2 \Kpm + n_3 \piz + n_4 \KS$ 
($n_1=1,...5$, $n_2=0,...2$, $n_3=0, ...2$ and $n_4=0,1$).
The $D^{\ast 0}$ is reconstructed in the decay mode
$D^0\piz$ and the $D^0$ candidate is reconstructed
in four decay modes:
$D^0 \to K^-\pip,K^-\pip\piz,K^-\pip\pim\pip,K^0_S\pim\pip$.

The selection of the fully reconstructed 
\B\ candidates is made according to the values of
two variables:
\begin{equation}
 \DeltaE = E_B^{\ast}-E_{beam}\;,
\end{equation}
where $E_B^{\ast}$ is the energy of the \B\ meson and $E_{beam}$ is
the beam energy, both in the \FourS\ rest frame;
\mes, the energy substituted mass, defined as:
\begin{equation}
\mes =  \sqrt{[ (s/2+{\bf p}
      \cdot{\bf p_\B})^2 / E^2 ]
  - |{\bf p_\B}|^2} \; ,
\end{equation}
where $\sqrt{s}$ is the total energy of the \epem\ system in the
\FourS\ rest frame, and $(E, {\bf p})$ and $(E_\B, {\bf p_\B})$ 
are the four-momenta of the \epem\ system and the reconstructed 
\B\ candidate respectively, both in the laboratory frame. 
We require $-0.1<\DeltaE<0.08\gev$ and $\mes>5.21\gevcc$.

\begin{figure}[!htb]
  \begin{center}
    \includegraphics[height=6cm]{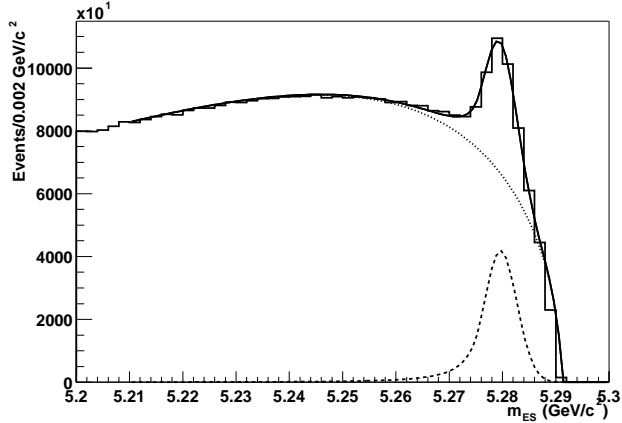}
    \caption{Distribution of the energy substituted mass \mes\ in
      data for the fully reconstructed \B\ mesons (histogram).
      The solid curve shows the result of the fit (see text).
      Also shown are the signal (dashed curve) and the background
      (dotted curve) components.} 
    \label{fig:dat_nopresel}
  \end{center}
\end{figure}

For each reconstructed \Breco\ mode $i$ 
the \mes\ distribution of the reconstructed \B\ candidates
is fit with the sum of an Argus function~\cite{ref:argus} and
a Crystal Ball function~\cite{ref:cb}. The Argus function models 
the continuum and combinatorial background whereas the Crystal Ball
models the signal component, which peaks at the \B\ mass.
The purity of the mode $i$ is defined as $S_i/(S_i+B_i)$,
where $S_i$ ($B_i$) is the number of signal (background) events
with $\mes>5.27\gevcc$, as determined by the fit. 
In events with more than one reconstructed charged \B\
candidate we select the candidate reconstructed in the mode 
with the highest purity.
Figure~\ref{fig:dat_nopresel} shows the \mes\ distribution 
for all \Breco\ candidates in data.
The yield $N_{\BpBm}$ of the sample containing one \Breco\ 
is determined as the area of the fitted Crystal Ball function. 
We obtain $N_{\BpBm} = (1.67 \pm 0.09) \times 10^5$.
The error on  $N_{\BpBm}$ is dominated by systematics and is
discussed in Section~\ref{sec:nbberror}.

We define the signal region on the \Breco\ side to be
$-0.09<\DeltaE<0.06\gev$ and $\mes>5.27 \gevcc$
and we use the events contained in the sideband $5.21 < \mes < 5.26
\gevcc$ as a control sample for continuum and combinatorial background.
The same \Breco\ reconstruction technique has been used in other
\babar\ analyses like~\cite{ref:knn}.

\subsection{Selection of {\boldmath \Btaunu} decays}
\label{sec:selections}

In the events where a \Breco\ is reconstructed
we search for decays of the recoil \B\
in a $\tau$ plus a neutrino;
the $\tau$ lepton is identified in the following decay channels:
\taumtoe, \taumtomu, \taumtopi, $\taum \to \pim \piz \nu$, $\taum
\to \pim \pip \pim \nu$.
All the selection criteria have been optimized to achieve the 
best upper limit.

The possible modes in which a \Breco\ meson can be reconstructed 
have been classified by decreasing purity. For each reconstructed
$\tau$ decay channel we select only \Breco\ mesons reconstructed
in the first $n$ modes, where $n$ has been chosen
for the best upper limit.

The event total charge $q$ is defined as the sum of the \Breco\ 
charge plus the $\tau$ decay products charge.
We consider only events with total charge $q=0$ (right sign sample).
The complementary sample with $|q|=2$ (wrong sign sample) contains
a negligible fraction of the signal and is used as a control sample to
test the analysis strategy and the agreement between the
selected events in data and the expectation from Monte Carlo
simulations.

All the physical quantities mentioned in the following, except 
where explicitly stated, refer to the recoil \B.

\subsubsection{Selection of \boldmath \taumtoe, \taumtomu and \taumtopi decays}
The \taumtoe, \taumtomu and \taumtopi\ channels are
characterized by a single charged track in the final
state coming from the primary vertex.

We require:
\begin{itemize}
\item one reconstructed charged track which has not been
      identified as a kaon, 
      no reconstructed \piz\ and  no reconstructed \KS;
\item at most one photon candidate with total energy in the laboratory
  frame less than 110~\mev. Only photons of at least 50~\mev are 
  considered; 
\item at least 1.2~\gevc of missing momentum in the laboratory frame;
\item the track must be identified as a lepton for the
  \taumtoe\ and \taumtomu\ selections;
\item for the \taumtopi\ selection we require the charged track to have 
  a momentum in the recoil \B\ rest frame of at least 1.2~\gevc and 
  not be identified as either an electron or a muon.
\end{itemize}

\subsubsection{Selection of \boldmath $\taum \to \pim \piz \nu$ decay}
The $\taum \to \pim \piz \nu$ decay proceeds via an intermediate
$\rho^-$ state. 
We require:
\begin{itemize}
\item one reconstructed track which has not been identified as 
  a kaon or a lepton, one reconstructed \piz and no reconstructed \KS;
\item at least 1.4~\gevc\ of missing momentum in the laboratory frame;
\item at most one photon candidate with total energy in the laboratory
  frame less than 100~\mev.
  Only photons of at least 50~\mev and not used for the \piz\
  reconstruction are considered; 
\item the invariant mass of the $\pim \piz$ pair has to be
  in the range $0.55 < m_{\pim \piz}< 1.0 \gevcc$;
\item for a further rejection of  the continuum background, we require 
  the cosine of the angle between the direction of the momentum
  of the \Breco\ and the thrust vector of the recoil \B\
  to be less than 0.9; the thrust orientation is chosen in order to 
  point in the hemisphere opposite to the direction of the recoil \B\ momentum.
\end{itemize}

\subsubsection{Selection of \boldmath $\taum \to \pim \pip \pim \nu$ decay}
The $\taum$ decays into three charged tracks via two
intermediate resonances. The full decay chain is: 
$\taum\to a^-_1 \nu$, $a^-_1 \to \pim \rho^0$, $\rho^0 \to \pip \pim$.
We require:
\begin{itemize}
\item three reconstructed charged tracks which have not been 
  identified as leptons or kaons, no reconstructed $\piz$ and
  no reconstructed \KS;
\item at least 1.2~\gevc of missing momentum in the laboratory frame;
\item at most one photon candidate with total energy in the laboratory
  frame less than 100~\mev.
  Only photons of at least 50~\mev and satisfying the quality requirements 
  on the lateral moment~\cite{ref:lat} to be between $0.05$ and $0.50$
  and on $\Sigma_9 / \Sigma_{25} > 0.9$ are considered.
  The lateral moment is a shape quantity for a neutral cluster and 
  $\Sigma_9 / \Sigma_{25}$ is the ratio of the energies deposited in the 9 
  and 25 crystals closest to the cluster centroid. These quality
  requirements are introduced to improve the description of the neutral
  energy distribution obtained from Monte Carlo simulations;
\item at least one $\pim \pip$ pair with invariant mass in the range 
  $0.60<m_{\pip\pim}<0.95 \gevcc$;
\item invariant mass of the three pions in the range 
  $1.1<m_{\pim\pip\pim}<1.6 \gevcc$;
\item total momentum of the three pions in the recoil \B\ rest frame
  greater than 1.6~\gevc.
\end{itemize}

\subsection{Efficiency and expected background}
\label{sec:effbkg}
The selection efficiencies for the $\tau$ decay channels we
consider in this analysis are determined from detailed Monte
Carlo simulations and are summarized in Table~\ref{tab:eff}. 
We compute the efficiency as the ratio of the number of events
surviving each of our selections and the number of events
where a \Breco\ has been reconstructed.
The efficiency for the \taumtomu\
channel is three times lower than the efficiency for the \taumtoe\ channel
because a large fraction of the muon momentum spectrum is below $1\gevc$,
where the muon selection efficiency is low. These muons are not recovered
by the pion selection because we require the pion momentum to be at least
$1.2\gevc$ in order to reject combinatorial and continuum backgrounds.

In the computation of the total efficiency for each selection 
we have taken into account the cross-feed from the other $\tau$ decay
channels reported in Table~\ref{tab:eff},
the requirement that the \Breco\ is reconstructed 
in the signal region and that the total reconstructed event charge is zero.

\begin{table}[!htb]
  \caption{Efficiency of the different selections (columns) for the 
    most abundant $\tau$ decay channels (rows).
    In case the efficiency is zero we quote a 90\% C.L. upper limit. 
    The last two rows show the total efficiency of the
    single selections, weighted by the decay branching fractions, and
    the total efficiency. The errors are statistical only.
    The total efficiency for each selection is:
    $\epsilon_i = \sum_{j=1}^{n_{dec}} \epsilon_i^j f_j$,
    where $\epsilon_i^j$ is the efficiency of the selection $i$
    for the $\tau$ decay channel $j$, $n_{dec}=7$ is the number of rows in
    the table and $f_j ={\cal
    B}(\tau\to j)$ are the 
    $\tau$ branching fractions from Ref.~\cite{ref:pdg2002}.}
  \begin{center}
    \begin{tabular}{|l|c|c|c|c|c|} \hline
 mode                & $e\nu\nu$ (\%)  & $\mu\nu\nu$ (\%) & $\pi\nu$ (\%)    & $\pim\pip\pim\nu$ (\%)& $\pim\piz\nu$ (\%) \\ \hline
$e\nu\nu$        & {\bf 22.9 $\pm$ 0.6}&  0 ($<$0.09)     &  0.1 $\pm$ 0.1   &  0 ($<$0.09)          &  0 ($<$0.09)         \\
$\mu\nu\nu$          &  0 ($<$0.08)    &{\bf 7.4$\pm$0.4} &  2.7 $\pm$ 0.2   &  0 ($<$0.08)          &  0.3 $\pm$ 0.1     \\
$\pi\nu$             &  0.6 $\pm$ 0.1  &  0.5 $\pm$ 0.1   &{\bf 21.6$\pm$0.6}&  0 ($<$0.11)          &  1.0 $\pm$ 0.2     \\
$\pim\pip\pim\nu$    &  0 ($<$0.15)    &  0 ($<$0.15)     &  0.4 $\pm$ 0.1   &{\bf 6.8$\pm$0.6}      &  0.1 $\pm$ 0.1     \\
$\pim\piz\nu$        &  0 ($<$0.05)    &  0 ($<$0.05)     &  1.2 $\pm$ 0.1   &  0 ($<$0.05)          &{\bf 6.6$\pm$0.3}   \\
$\pim\piz\piz\nu$    &  0 ($<$0.14)    &  0 ($<$0.14)     &  0 ($<$0.14)     &  0 ($<$0.14)          &  0.8 $\pm$ 0.2     \\
$\pim\pip\pim\piz\nu$&  0 ($<$0.03)    &  0 ($<$0.03)     &  0.1 $\pm$ 0.1   &  0 ($<$0.03)          &  0.6 $\pm$ 0.2     \\
\hline
 all $\tau$ dec.:    &{\bf 4.2 $\pm$ 0.1}&{\bf 1.3 $\pm$ 0.1}&{\bf 3.2 $\pm$ 0.1}&{\bf 0.6 $\pm$ 0.1}&{\bf 2.0 $\pm$ 0.1} \\ \hline
 total:              & {\bf 11.3 $\pm$ 0.2} & & & &\\ \hline
\end{tabular}

  \end{center}
  \label{tab:eff}
\end{table}

The expected background is determined separately in the 
right sign and wrong sign samples.
It is composed of events from continuum and combinatorial background, 
and events with a correctly reconstructed \B\ meson.
Simulations of \BzBzb\ events have shown that
events where a neutral \B\ is incorrectly reconstructed as a charged
\B\ provide a negligible peaking component.

The continuum and combinatorial background is determined from the 
number of events in the \mes\ sideband, scaled by the ratio of the areas 
of the fitted Argus function in the signal and sideband regions. 
Since the number of background events after the full selection is
too small to perform a precise fit, we define for each selection
criterion a preselection based on the requirements on the number
of reconstructed charged tracks and \piz\ mentioned in
section~\ref{sec:selections}.
We  fit the \mes\ distribution after each preselection and we assume 
that the ratio of the fitted Argus in sideband and signal regions,
which we use in our estimate of the continuum and combinatorial
background, is unchanged after the full selection.
The peaking background is determined from Monte Carlo simulations
of \BpBm\ events. 

Another source of background originates from \epem\to\tautau\ events.
From Monte Carlo simulations we expect $5.8\pm 1.9$ events 
from \tautau\ that survive the \taumtopi\ selection. No \tautau\ 
event survives in the wrong sign sample.
The expected background is summarized in Tables~\ref{tab:bkg-ws}
and~\ref{tab:bkg-rs} for the wrong sign and right sign samples,
respectively. The systematic corrections on the expected background
are described in the next Section.

\begin{table}[!htb]
  \caption{Expected background for the wrong sign sample.
    The peaking component is estimated from inclusive
    \BpBm\ Monte Carlo and the combinatorial plus continuum component
    from the data sideband. If no event survives the selection we quote
    a 90\% C.L. upper limit on the expected background.   
    Systematic corrections are not included.} 
  \begin{center}
\begin{tabular}{|l|c|c|c|} \hline
 selection         & peaking        & cont. + comb.  & total bkg.      \\ \hline
 $e\nu\nu$         &  4.6 $\pm$ 1.8 &  0.6 $\pm$ 0.4 &  5.2 $\pm$ 1.8  \\ 
 $\mu\nu\nu$       &  0.7 $\pm$ 0.7 &  0 ($<$1.4)    &  0.7 $\pm$ 0.7  \\ 
 $\pi\nu$          &  5.3 $\pm$ 1.9 &  0 ($<$1.4)    &  5.3 $\pm$ 1.9  \\ 
 $\pim\pip\pim\nu$ &  2.0 $\pm$ 1.2 &  2.1 $\pm$ 0.8 &  4.1 $\pm$ 1.4  \\ 
 $\pim\piz\nu$     &  9.4 $\pm$ 2.6 &  0 ($<$1.4)    &  9.4 $\pm$ 2.6  \\ 
\hline
 all &  &  &24.7 $\pm$ 4.0 \\ \hline
\end{tabular}

\end{center}
\label{tab:bkg-ws}
\end{table}

\begin{table}[!htb]
  \caption{Expected background for the right sign sample.
    The peaking component is estimated from inclusive \BpBm\
    Monte Carlo and the combinatorial plus continuum from the
    data sideband. The contribution from the \tautau\ background
    is also shown. If no event survives the selection we quote
    a 90\% C.L. upper limit on the expected background.
    Systematic corrections are not included.}
  \begin{center}
\begin{tabular}{|l|c|c|c|c|} \hline
 selection         & peaking        & cont. + comb.  & \tautau\ bkg.  & total bkg.\\ \hline
 $e\nu\nu$         &  7.2 $\pm$ 2.1 &  0 ($<$1.4)    &      0 ($<$1.5)&  7.2 $\pm$ 2.1  \\ 
 $\mu\nu\nu$       &  4.7 $\pm$ 1.8 &  0.6 $\pm$ 0.4 &      0 ($<$1.5)&  5.3 $\pm$ 1.8  \\ 
 $\pi\nu$          &  4.3 $\pm$ 1.6 &  1.3 $\pm$ 0.6 &  5.8 $\pm$1.9  & 11.4 $\pm$ 2.5  \\ 
 $\pim\pip\pim\nu$ &  1.6 $\pm$ 1.1 &  3.0 $\pm$ 1.0 &      0 ($<$1.5)&  4.6 $\pm$ 1.5  \\ 
 $\pim\piz\nu$     & 10.3 $\pm$ 2.9 &  1.7 $\pm$ 0.7 &      0 ($<$1.5)& 12.0 $\pm$ 3.0  \\ 
\hline
 all              &                 &                &                & 40.5 $\pm$ 5.0  \\ 
\hline
\end{tabular}

\end{center}
\label{tab:bkg-rs}
\end{table}

\begin{figure}[!htbp]
  \begin{center}
    \begin{tabular}{cc}
      \includegraphics[height=5cm]{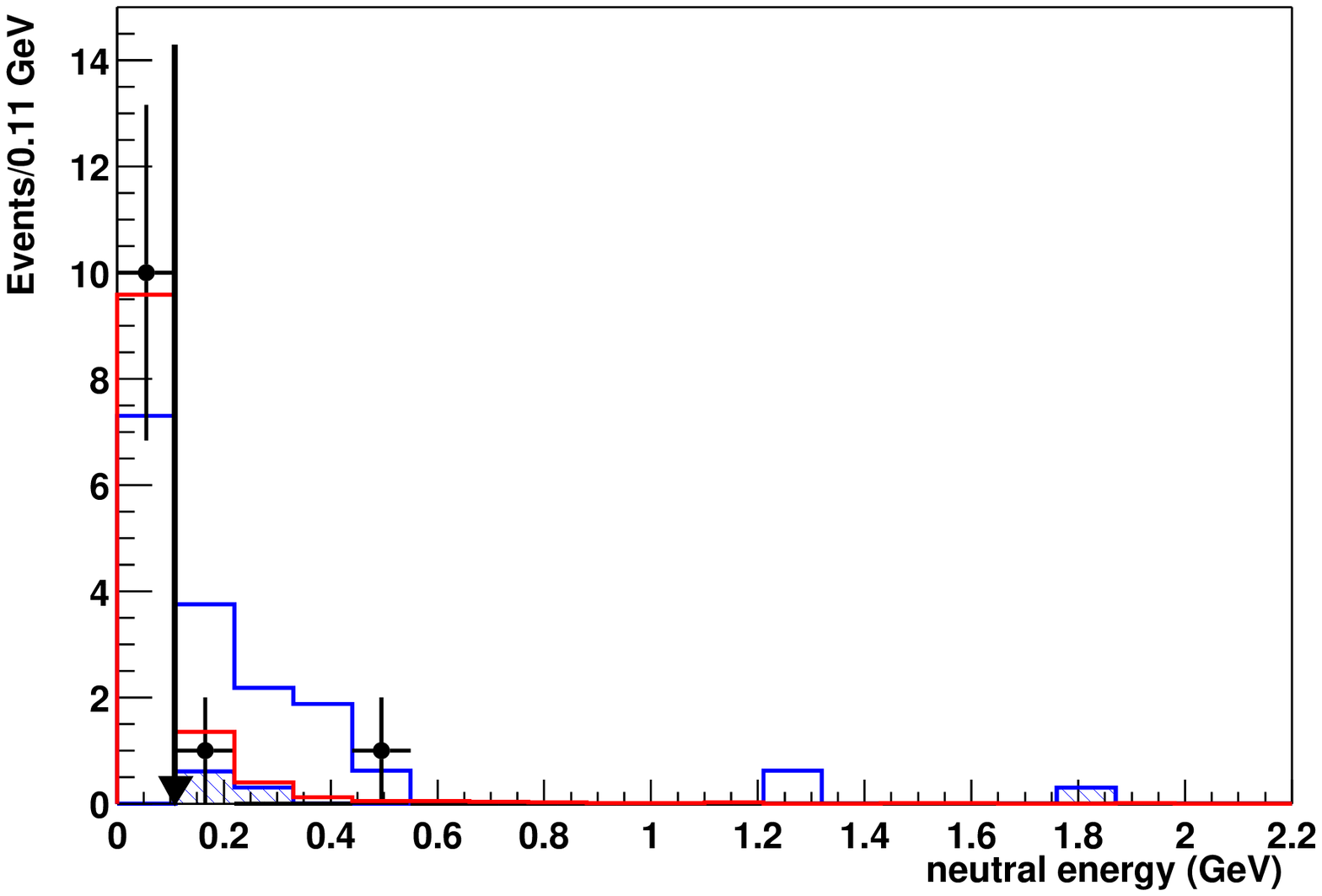} &
      \includegraphics[height=5cm]{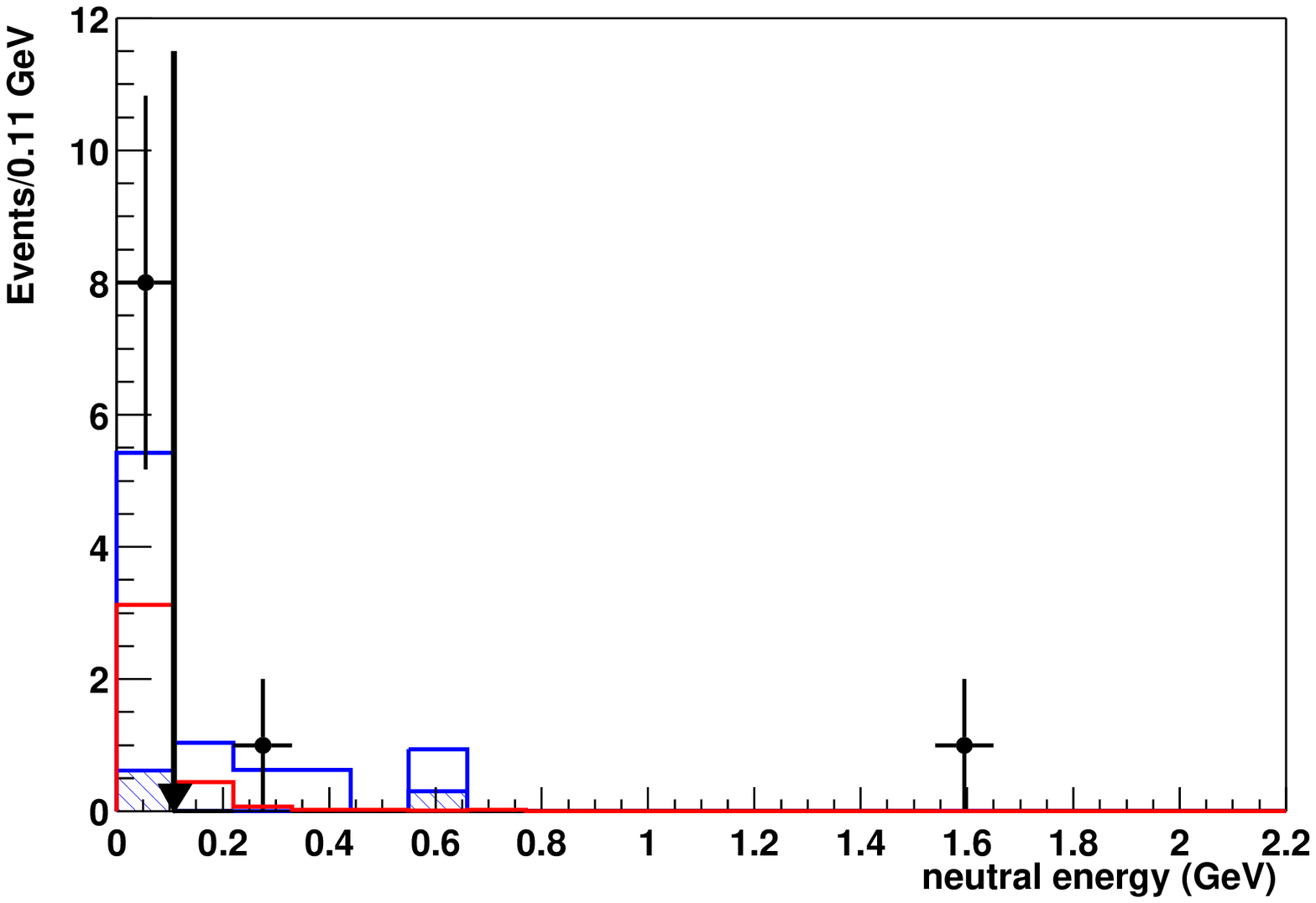} \\
      (a) & (b) \\
    \end{tabular}
    \begin{tabular}{cc}
      \includegraphics[height=5cm]{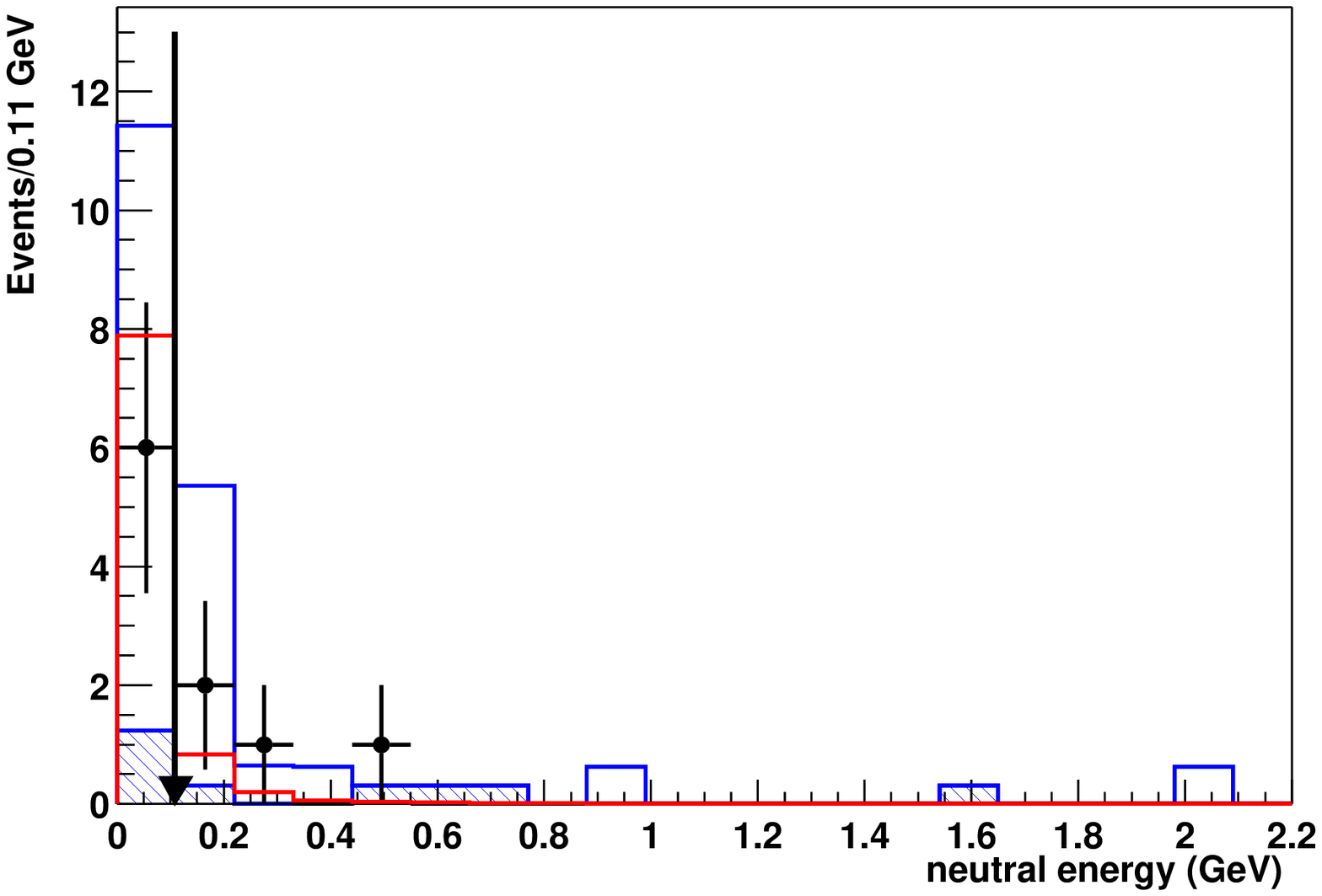} &
      \includegraphics[height=5cm]{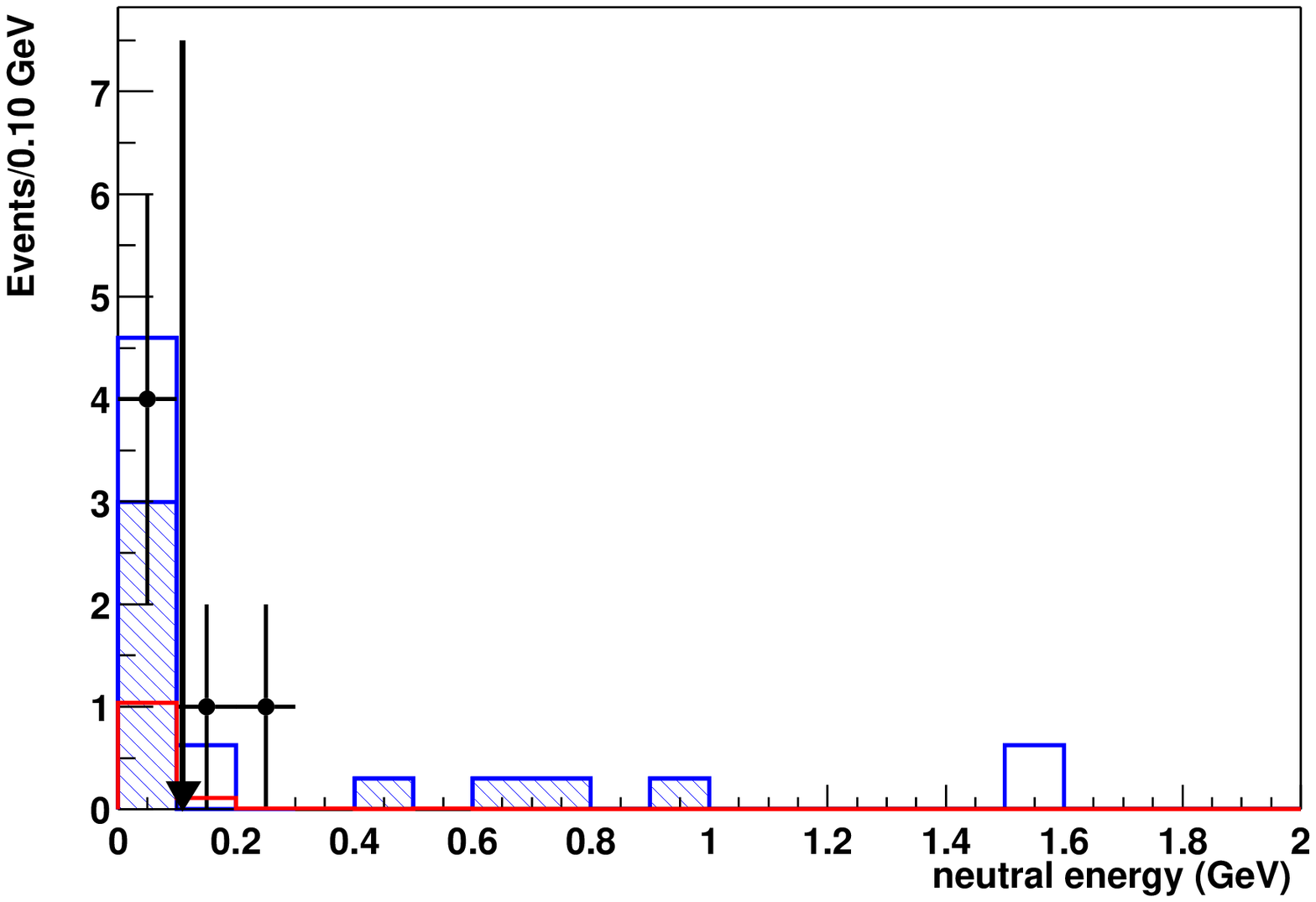} \\
      (c) & (d) \\      
    \end{tabular}
    \begin{tabular}{c}
      \includegraphics[height=5cm]{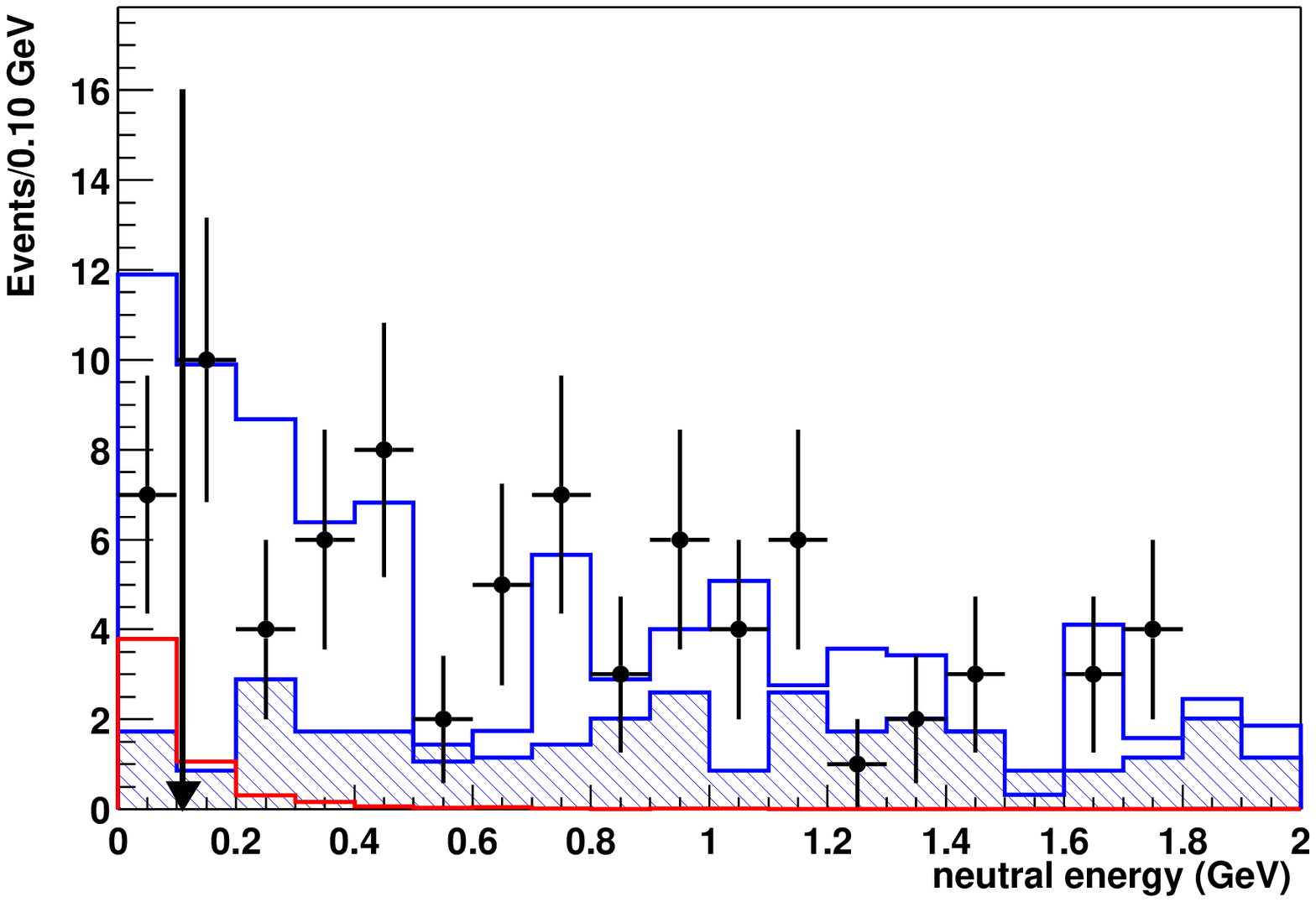} \\
        (e) \\      
    \end{tabular}
    \caption{Neutral energy distribution in the laboratory frame after all
    the selection requirements except the one on the neutral energy for
    the channels (a)\taumtoe, (b)\taumtomu, (c)\taumtopi, 
    (d)$\tau\to\pim\pip\pim\nu$ and (e)$\tau\to\pim\piz\nu$.
    The shaded histogram is the continuum plus combinatorial component 
    of the expected background; the solid histogram represents the peaking
    component of the expected background and the background from \tautau\ events;
    the dots are the data in the \mes\ signal region; the red light shaded 
    histogram represents the distribution for Monte Carlo simulated
    signal events scaled to \BR(\Btaunu)=$10^{-3}$. The vertical arrow is
    the requirement on the neutral energy in each selection.}
    \label{fig:final}
  \end{center}
\end{figure}

In Fig.~\ref{fig:final} we show the neutral energy distribution for
events in data and for the expected background. Each distribution
refers to a different selection and is obtained applying all the 
requirements except the one on the neutral energy. 
The plots show no evidence of signal in data.

\section{Systematic uncertainties}
\label{sec:Systematics}
The main sources of uncertainty in the determination of the \Btaunu\ branching
fraction are:
\begin{itemize}
\item uncertainty in the determination of the efficiency $\epsilon_i$
  for each selection channel;
\item uncertainty in the determination of the number of \BpBm\ events
with one reconstructed \Breco\, $N_{\BpBm}$;
\item uncertainty in the determination of the number of expected
  background events $b_i$ in each selection channel.
\end{itemize}

\subsection{\boldmath Uncertainty in the selection efficiencies}
The main contributions to the systematic uncertainties in the
determination of the efficiencies come from systematic uncertainty on
tracking efficiency, neutral reconstruction 
efficiency and particle identification (PID).
Uncertainty in the \piz\ reconstruction efficiency introduces 
an additional 5\% contribution to the systematics in the
$\taum \to \pim \piz \nu$ selection .
The different contributions to the systematic 
uncertainties on the selection efficiencies are 
reported in Table~\ref{tab:syseff}.

\begin{table}[!htb]
  \caption{Contributions to the systematic uncertainty
        on the efficiency of the different selections. In the $\piz\nu$
        channel the contribution to the neutral systematic uncertainty
        due to \piz\ reconstruction is reported explicitly.}
  \begin{center}
  \begin{tabular}{|l|c|c|c|c|} \hline
 selection        & tracking (\%) & PID (\%)& neutral reco (\%) & total (\%)\\ \hline
 $e\nu\nu$        &  0.8          & 2.4     &  0.9              & 2.7       \\
 $\mu\nu\nu$      &  0.8          & 6.0     &  0.9              & 6.1       \\
 $\pi\nu$         &  0.8          & 6.0     &  0.9              & 6.1       \\
 $\pim\pip\pim\nu$&  2.4          &11.4     &  3.8              &13.6       \\
 $\pim\piz\nu$    &  0.8          & 3.2     &  1.1$\oplus$5     & 6.1       \\
 \hline
\end{tabular}

  \end{center}
  \label{tab:syseff}
\end{table}

\subsection{\boldmath Uncertainty in the determination 
        of $N_{\BpBm}$ }
\label{sec:nbberror}
We determine $N_{\BpBm}$ as the area of the Crystal Ball function 
fitted to the \mes\ distribution (see Fig.~\ref{fig:dat_nopresel}).
Using a Gaussian function as an alternative description of 
the peak, we obtain a value of $N_{\BpBm}$ which is smaller by 4.5\%. 
We assume this relative difference as the 
systematic uncertainty on $N_{\BpBm}$. 
Using the product of a third order polynomial times an Argus 
function as an alternative model for the background,
the change in $N_{\Bp\Bm}$ is 0.6\%.

\subsection{\boldmath Uncertainty in the 
expected background and systematic corrections}

To take into account possible dependencies of the 
fitted Argus shape on a given variable used in the
selections, we compute a correction factor as the ratio 
of the expected background events passing the requirement
on it using two different approaches.
In the first approach we use a single sideband to signal
scaling factor (see Section~\ref{sec:effbkg}) determined
from a \mes fit over the full variable range.
In the second approach we divide the range of the variable
into bins and determine different scaling factors for
each bin.
To each correction factor we assign 100\% of the deviation
from unity as a systematic uncertainty.

The expected number of background events after the correction
is shown in Tables~\ref{tab:bkgcor-ws} and~\ref{tab:bkgcor-rs}
for the wrong sign and right sign samples, respectively.
It agrees with the number of selected events in data.
The total systematic uncertainty amounts
to  8.3\% for the \taumtomu\ and \taumtoe\
channels, 9.4\% for the \taumtopi\ channel, 9.9\% for the 
$\taum \to \pim \piz \nu$ channel, and 6.1\% for the 
$\taum \to \pim \pip \pim \nu$ channel.

\begin{table}[!htb]
  \caption{Corrected expected background for the wrong sign sample
    compared to the number of the selected 
    data candidates. The errors are the statistical and systematic
    uncertainties.} 
  \begin{center}
    \begin{tabular}{|l|c|c|} \hline
 selection & corr. total bkg. & data candidates  \\ \hline
 $e\nu\nu$ &  4.9 $\pm$ 1.7 $\pm$ 0.4 &    5 \\ 
 $\mu\nu\nu$ &  0.6 $\pm$ 0.6 $\pm$ 0.1 &    3 \\ 
 $\pi\nu$ &  5.2 $\pm$ 1.8 $\pm$ 0.5 &    0 \\ 
 $\pim\pip\pim\nu$ &  3.8 $\pm$ 1.3 $\pm$ 0.3 &    3 \\ 
 $\pim\piz\nu$ &  8.1 $\pm$ 2.3 $\pm$ 0.9 &    9 \\ 
\hline
 all & 22.7 $\pm$ 3.7 $\pm$ 1.2 &   20 \\ \hline
\end{tabular}

  \end{center}
\label{tab:bkgcor-ws}
\end{table}

\begin{table}[!htb]
  \caption{Corrected expected  background for the right sign sample
   compared to the number of the selected data candidates.
   The errors are the statistical and systematic
    uncertainties.}
  \begin{center}
    \begin{tabular}{|l|c|c|c|} \hline
 selection         & corr. total bkg.         &  data candidates  & exp. signal events\\  & & & for $\BR(\Btaunu) = 10^{-4}$  \\ \hline
 $e\nu\nu$         &  6.7 $\pm$ 2.0 $\pm$ 0.6 &    10             &   0.7 \\ 
 $\mu\nu\nu$       &  5.0 $\pm$ 1.7 $\pm$ 0.4 &     8             &   0.2 \\ 
 $\pi\nu$          & 11.2 $\pm$ 2.5 $\pm$ 0.5 &     6             &   0.5 \\ 
 $\pim\pip\pim\nu$ &  4.3 $\pm$ 1.4 $\pm$ 0.3 &     4             &   0.1 \\ 
 $\pim\piz\nu$     & 10.4 $\pm$ 2.6 $\pm$ 1.0 &     7             &   0.3 \\ 
\hline
 all               & 37.6 $\pm$ 4.7 $\pm$ 1.3 &    35             &   1.8 \\ \hline
\end{tabular}

  \end{center}
\label{tab:bkgcor-rs}
\end{table}

\section{Upper limit extraction}
In order to extract the upper limit on the branching fraction
for \Btaunu\ we combine the results of the different selections.

We use the likelihood ratio estimator:
\begin{equation}
  Q=\frac{{\cal L}(s+b)}{{\cal L}(b)} \: ,
\end{equation}
where ${\cal L}(s+b)$ and ${\cal L}(b)$ are the likelihood functions
for signal plus background and background only hypotheses, respectively. 
The likelihood functions ${\cal L}(s+b)$ and ${\cal L}(b)$ are defined
as:
\begin{eqnarray}
  {\cal L}(s+b) & = &
  \prod_{i=1}^{n_{ch}}\frac{e^{-(s_i+b_i)}(s_i+b_i)^{n_i}}{n_i!} \: ,
  \label{eq:lsb}
  \\
  {\cal L}(b) & = &
  \prod_{i=1}^{n_{ch}}\frac{e^{-b_i}b_i^{n_i}}{n_i!}  \: ,
  \label{eq:lb}
\end{eqnarray}
where $n_{ch}$ is the number of selection channels, $s_i$ and $b_i$
are the expected number of signal and background
events respectively and $n_i$ is the number of selected events
in each channel.
In particular, $s_i$ can be written in terms of \BRBtaunu\ as:
\begin{equation}
  s_i = s \epsilon_i = 
  N_{\Bp\Bm}\BRBtaunu\epsilon_i\ ,
\end{equation}
where $s$ is the total expected number of \Btaunu\ events,
$\epsilon_i$ is the selection efficiency for the $i$-th channel, 
$N_{\Bp\Bm}$ is the number of \BpBm\ events with one 
reconstructed \Breco.

We have no evidence of signal
and we set a $90\%$ C.L. upper limit using a fast 
parametric Monte Carlo
generating random experiments for different values of the
branching fraction \BRBtaunu. The confidence level for the signal
hypothesis can be computed as:
\begin{equation}
  {\rm C.L.}_s = \frac{{\rm C.L.}_{s+b}}{{\rm C.L.}_b} = 
  \frac{N_{Q_{s+b}\le Q}}{N_{Q_b\le Q}} \: ,
\end{equation}
where $N_{Q_{s+b}\le Q}$ and $N_{Q_b\le Q}$ are the number of
the generated experiments which have a likelihood ratio less than
or equal to the measured one, in the background plus signal and
background only hypothesis respectively. 
The 90\% C.L. upper limit to the branching fraction is the value
for which ${\rm C.L.}_s=1-0.9$. We determine:
\begin{equation}
  \BRBtaunu < 6.3 \times 10^{-4}\:,\:\: 90\%\:{\rm C.L.}
\end{equation}

In the extraction of the above limit we have included 
the uncertainty on the efficiency by reducing
the efficiencies by one standard deviation (adding in 
quadrature the statistical and systematic uncertainty), 
and we have assumed conservatively the estimate
of $N_{\Bp\Bm}$ obtained with a Gaussian model
instead of a Crystal Ball.

\begin{figure}[!htb]
  \begin{center}
    \begin{tabular}{cc}
    \includegraphics[height=5.5cm]{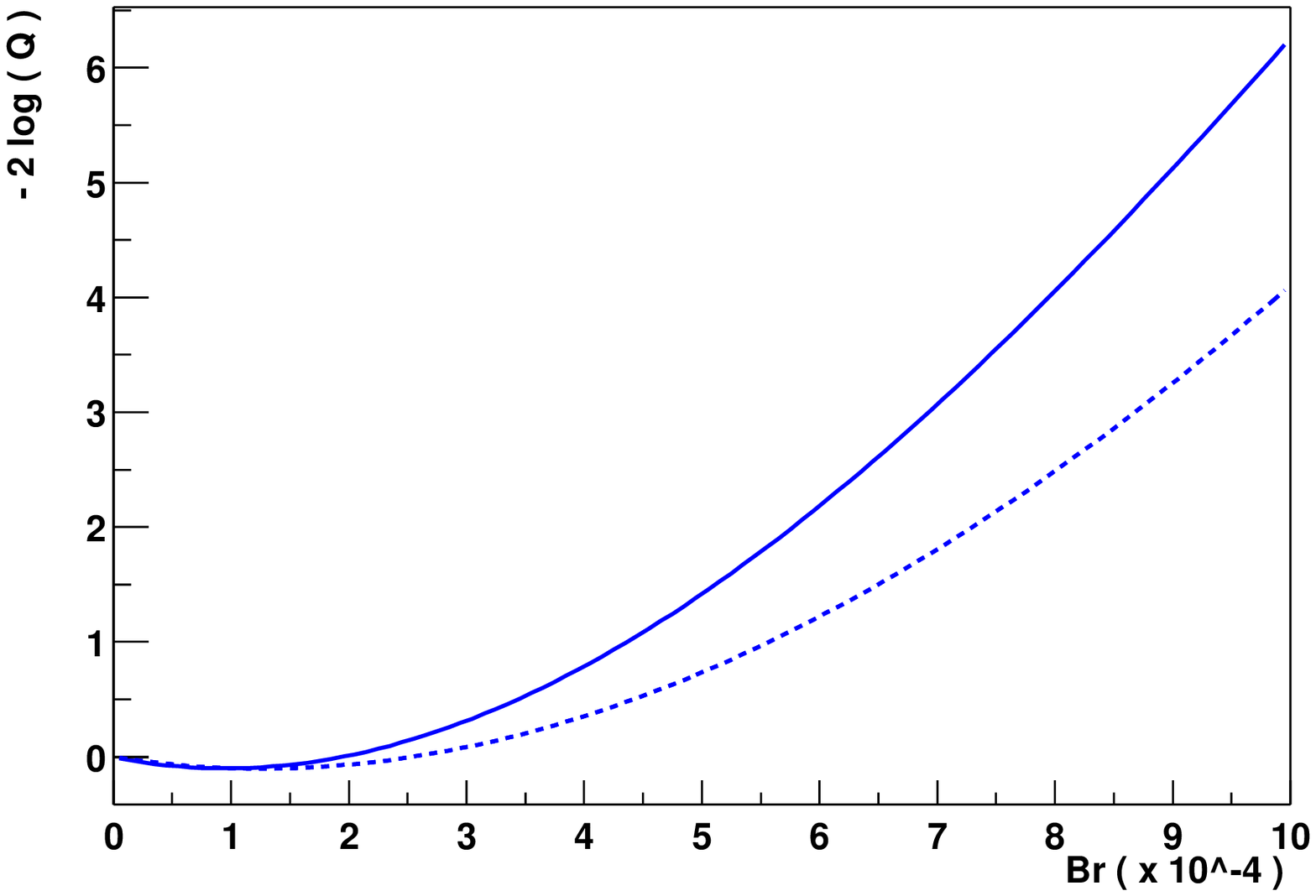} &
    \includegraphics[height=5.5cm]{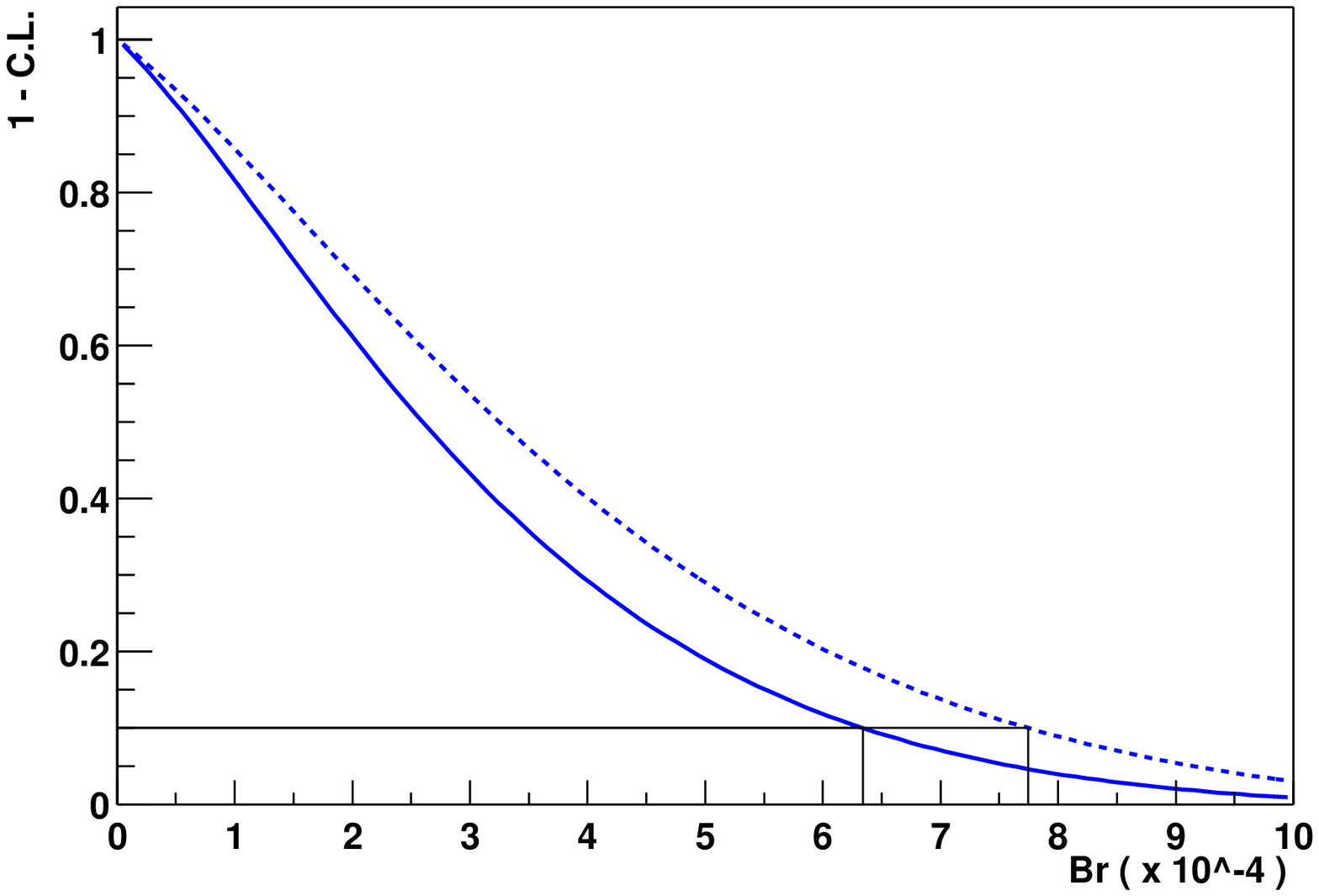} \\
    \end{tabular}
    \caption{Distributions of the likelihood ratio (left) and confidence level
        (right) as a function of \BRBtaunu. The dashed (solid) curve corresponds
        to the case in which the uncertainty on the expected background is included
        (not included).}
    \label{fig:clerr}
  \end{center}
\end{figure}

The statistical and systematic uncertainties on the expected 
background can be included in the likelihood definition by folding it 
with a Gaussian distribution 
having as standard deviation the combined statistical
and systematic error on the estimate of $b_i$. 
The effect of the uncertainty on the
expected background is shown in Fig.~\ref{fig:clerr}.
Including this uncertainty the upper limit becomes:
\begin{equation}
  \BRBtaunu < 7.7 \times 10^{-4}\:,\:\: 90\%\:{\rm C.L.}
\end{equation}

The central value of the branching fraction corresponds to
the minimum in the likelihood ratio distribution. Using
$N_{\Bp\Bm}$ obtained with a Crystal Ball model and the
central values of the efficiencies,
we determine $\BRBtaunu=(1.1^{+3.8}_{-1.1}\times 10^{-4})$.

\begin{figure}[!htb]
  \begin{center}
    \includegraphics[height=6cm]{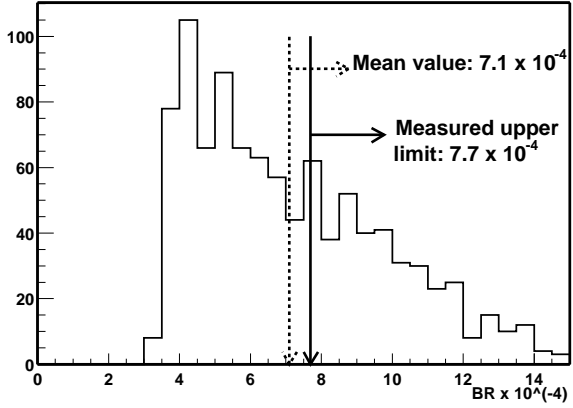}
    \caption{Distribution of the upper limit
      obtained by generating the selected data events according to 
      Poisson distributions. The statistical and systematic uncertainties
      on the expected background are taken into account.
      The dashed line indicates the nominal sensitivity, whereas the
      solid line shows the upper limit extracted from the data.}
    \label{fig:cldistrerr}
  \end{center}
\end{figure}

If we let the number of selected events in each channel
fluctuate according to Poisson distribution with 
the number of observed events as a mean
we obtain the distribution of the possible upper
limits shown in Fig.~\ref{fig:cldistrerr}.
The central value of this distribution ($7.1\times 10^{-4}$) 
represents our sensitivity to the upper limit.

The \babar\ Collaboration performed also another search for the 
\Btaunu\ decay using a statistically independent 
sample~\cite{ref:wisconsin}.
The sample is defined by one \Bp\ meson decaying in
$\bar{D}^0 \ell^+\nu_{\ell}X$ final state where $X$
is either a photon, \piz\ or nothing.
The two upper limits have been combined using the statistical
technique described above to combine several channels.
The combined upper limit is:

\begin{equation}
  \BRBtaunu < 4.1 \times 10^{-4}\:,\:\: 90\% \:{\rm C.L.}
\end{equation}

\section{Summary}
\label{sec:Summary}
A search for \Btaunu\ has been performed
in the recoil of a fully reconstructed \Breco\ sample.
The analysis uses the following $\tau$ decay channels: 
\taumtoe, \taumtomu, \taumtopi, $\taum \to \pim \piz \nu$
and $\taum \to \pim \pip \pim \nu$.
The results of the search in the different channels have been
combined using a likelihood approach.
No signal is observed and an upper limit has been set:
\[
  \BRBtaunu < 7.7 \times 10^{-4}\:,\:\: 90\%\: {\rm C.L.}
\]
The upper limits set by the two independent 
\Btaunu\ searches in the \babar\ experiment have been combined
using the statistical technique described in this paper to obtain
the following result:
\[
  \BRBtaunu < 4.1 \times 10^{-4}\:,\:\: 90\%\: {\rm C.L.}
\]
All results are preliminary.

\section{Acknowledgments}
\label{sec:Acknowledgments}

We are grateful for the 
extraordinary contributions of our \pep2\ colleagues in
achieving the excellent luminosity and machine conditions
that have made this work possible.
The success of this project also relies critically on the 
expertise and dedication of the computing organizations that 
support \babar.
The collaborating institutions wish to thank 
SLAC for its support and the kind hospitality extended to them. 
This work is supported by the
US Department of Energy
and National Science Foundation, the
Natural Sciences and Engineering Research Council (Canada),
Institute of High Energy Physics (China), the
Commissariat \`a l'Energie Atomique and
Institut National de Physique Nucl\'eaire et de Physique des Particules
(France), the
Bundesministerium f\"ur Bildung und Forschung and
Deutsche Forschungsgemeinschaft
(Germany), the
Istituto Nazionale di Fisica Nucleare (Italy),
the Foundation for Fundamental Research on Matter (The Netherlands),
the Research Council of Norway, the
Ministry of Science and Technology of the Russian Federation, and the
Particle Physics and Astronomy Research Council (United Kingdom). 
Individuals have received support from 
the A. P. Sloan Foundation, 
the Research Corporation,
and the Alexander von Humboldt Foundation.


\begin{thebibliography}{99}
\bibitem{ref:cb}
The Crystal Ball function is defined as:
  \[
  f(x) = N \cdot \left\{
  \begin{array}{ll}
    \exp (-\frac{(x - \overline {x})^2}{2 \; \sigma^2})  &;  
( \overline {x} - x)/\sigma >   \alpha \\
    A \times (B - \frac{x - \overline {x}}{\sigma})^{-n} &;  
( \overline {x} - x)/\sigma \leq \alpha \\
  \end{array} \right. 
\]
  where $A \equiv \left ( \frac{n}{|\alpha|} \right ) ^n 
\times \exp(-|\alpha|^2/2)$ and
  $B \equiv \frac{n}{|\alpha|} - |\alpha|$.
  $N$ is a normalization factor, $\overline x$ and $\sigma$ are the peak position and width
  of the Gaussian portion of the function, $\alpha$ is the point at which the 
  function changes to the power function and $n$ is the exponent of the power function.
  $A$ and $B$ are defined so that the function 
 and its first derivative are continuous at $\alpha$.
\end{thebibliography}
\end{document}